\documentclass[epj,nopacs]{svjour}
\usepackage{graphicx}
\usepackage[dvips]{color}
\usepackage[latin1]{inputenc}   % for accentuated letters
\begin{document}

\title{Bimodalities: a survey of experimental data and models}

%\classification{<Replace this text with PACS numbers; choose from this list:
%                \texttt{http://www.aip..org/pacs/index.html}>}
%\keywords      {nuclear reactions; phase transition; bimodality}

\author{O. Lopez\inst{1}
\and M. F. Rivet\inst{2}
}

\institute{Laboratoire de Physique Corpusculaire, IN2P3-CNRS/ENSICAEN
/Universit\'e, F-14050 Caen cedex, France
\and Institut de Physique Nucl\'eaire, IN2P3-CNRS, F-91406 Orsay cedex,
France
}

\date{\today}     %% seulement avec option draft
%\tableofcontents  %% seulement avec option draft
%\newpage %% seulement avec option draft

\abstract{
Bimodal distributions of some chosen variables measured in nuclear 
collisions were recently proposed as a non ambiguous signature of 
a first order phase transition in nuclei. This section presents a compilation 
of both theoretical and experimental studies on bimodalities 
performed so far, in relation with the 
liquid-gas phase transition in nuclear matter.
}

\maketitle

 After a formulation of
the theoretical bases of bimodality,  world-wide experimental results 
will be reviewed and discussed, as well as the occurrence of some
kind of bimodality in models. Finally
conclusions on the perspectives of such analyses in the near future 
and the possible connections to other proposed signals of the 
liquid-gas phase transition in nuclear matter will be given. 

\section{Theoretical bases}

\subsection{Definition}

Bimodality is a property of finite systems undergoing a first order 
phase transition~\cite{Hil63,Chom01,Lee02}. It is thus a generic feature which 
concerns not only Nuclear Physics but a broad domain of physics
such as Astrophysics, 
or Soft Matter Physics. Bimodality means that \textit{ the probability
distribution of an order parameter of the considered system at phase
transition exhibits two peaks separated by a minimum}. 
Indeed, if the system is in a pure 
phase, the order parameter distribution consists in one peak and can
be characterized by its mean value and its variance. 
By contrast, if the system is in the coexistence region, the 
distribution presents two peaks, well separated, whose properties are 
related to the two different phases of the system~\cite{Chom01}. 
Bimodality is then one of the signals associated to a first order phase 
transition~\cite{Lee02}, beside others such as scaling laws, critical 
exponents or negative heat capacities. 

In the following, the term ``bimodality''
will abbreviate ``the probability distribution of some variable, in a 
given region of the phase diagram of the system, is bimodal''.

\begin{figure}
\includegraphics[width=0.9\columnwidth]{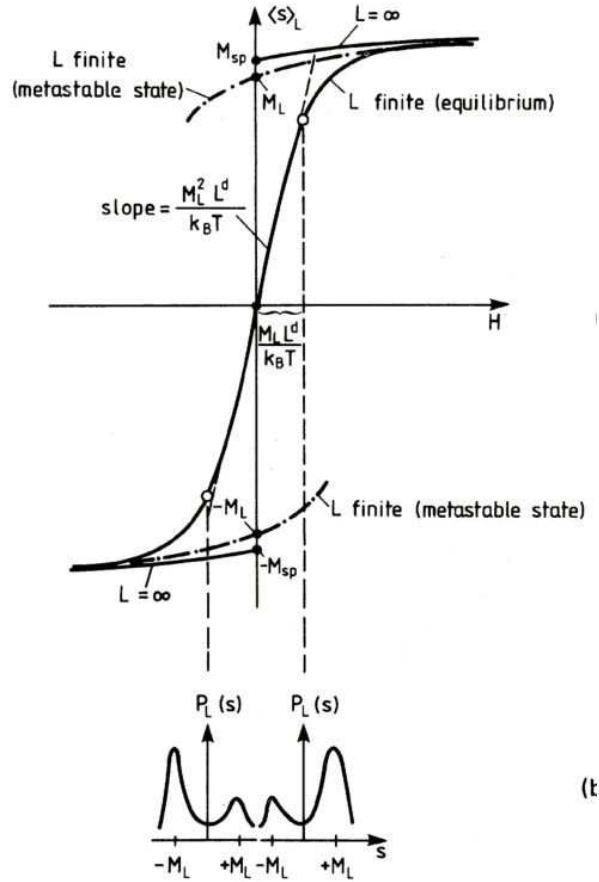}
\caption{Evolution of the magnetization $M$, as a function of the
applied  magnetic field $H$, in the Ising Model for a lattice defined
by the size L. The bottom panel presents a schematic  probability
distribution of the magnetization between $-M_L$ and $+M_L$ around the  
critical field value $H_{c}$.
Taken from~\cite{Bind84}.} \label{magnet}
\end{figure}

\begin{figure*}
\begin{center}
\includegraphics[scale=0.9]{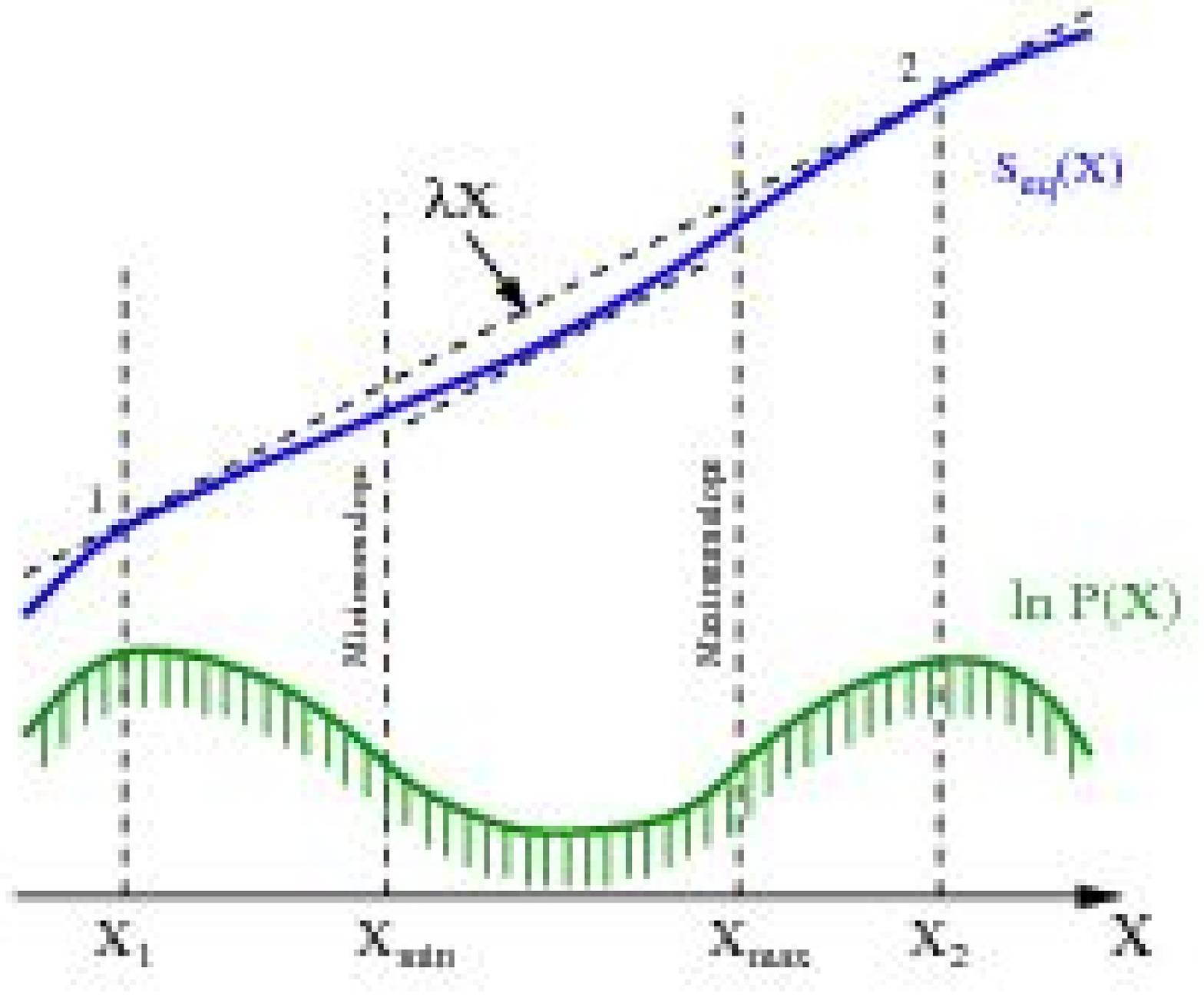}
\includegraphics[scale=0.9]{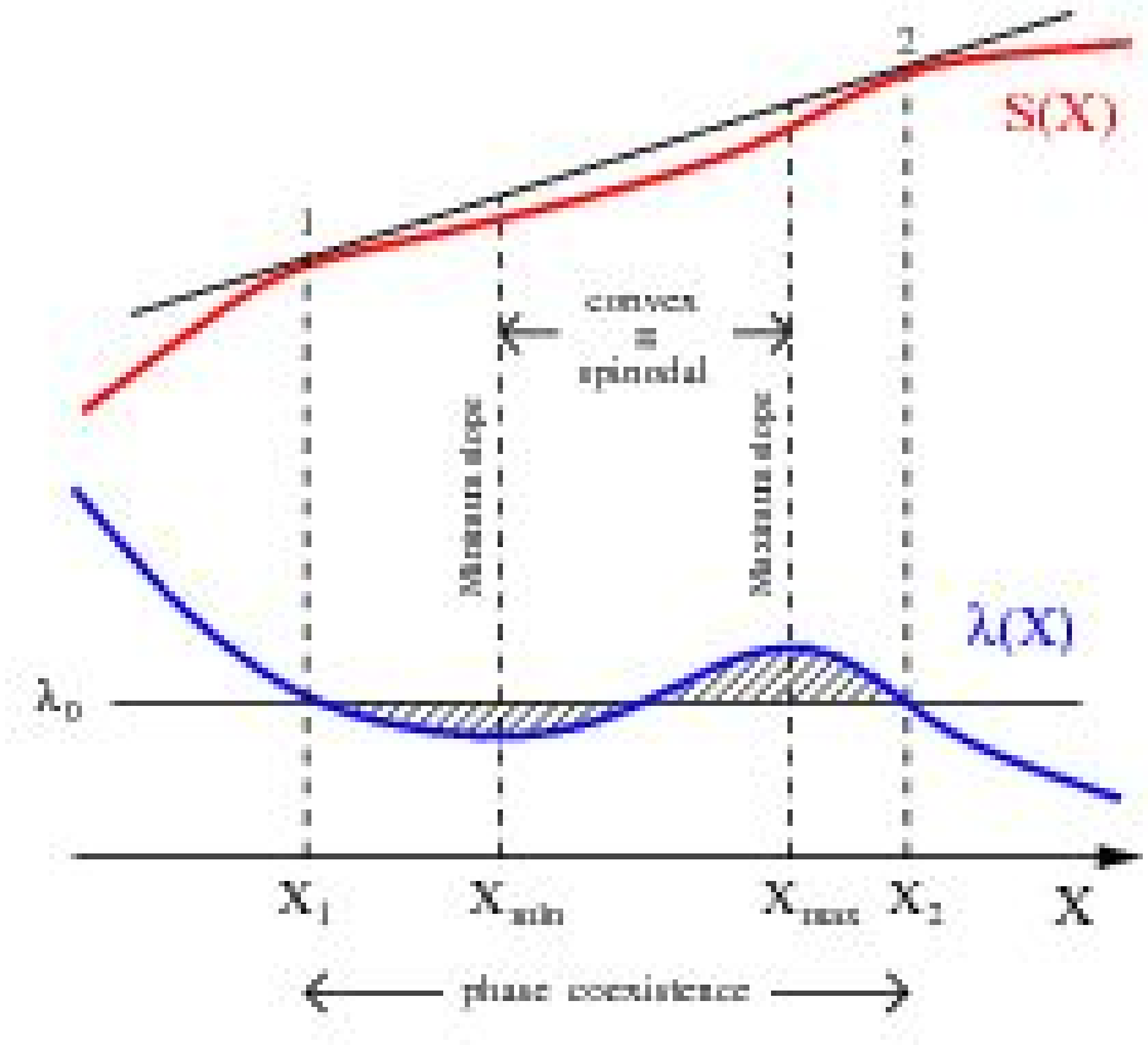}
\end{center}
\caption{Entropy \textit{S} of the system as a function of an order parameter 
\textit{X} of the phase transition. The relation is made between the convex 
 intruder of \textit{S}, the bimodal distribution in \textit{X} (left) and 
the abnormal  fluctuations of \textit{X} in  the phase coexistence region 
(right).  \textit{$\lambda$} is the intensive variable associated 
with \textit{X}.  Taken from~\cite{Chom04}.}\label{otherdef}
\end{figure*}

\subsection{Pioneering studies}
Bimodality and its relationship to phase transition has been studied 
since the 80's. Fig.~\ref{magnet} shows an Ising model simulation of a 
ferromagnet studied by Binder and Landau~\cite{Bind84}. In this analysis, 
the authors studied the magnetization $M$ of the system as a function 
of the applied magnetic field $H$. When the magnetic field comes close 
to the critical value $H_{c}$, the spontaneous magnetization of 
the ferromagnet presents a sudden change; 
in this case the probability distribution of the magnetization 
is never bimodal, as the system ``jumps'' suddenly from the negative 
value -$M_{sp}$ to the positive one +$M_{sp}$: the transition 
between the two regimes is sharp at the thermodynamical limit.

By contrast, when the size of the system is finite (and defined by the 
number of sites $L$), the step function is replaced by a smooth curve in 
fig.\ref{magnet}, with a slope proportional to $L^{d}$ - where $d$ is the 
dimensionality of the system. Consequently, in the vicinity of $H_{c}$, 
the magnetization $M$ exhibits a bimodal structure as shown in the 
bottom panel of fig.~\ref{magnet}. 

\subsection{Link with phase transition in Thermodynamics}
It was recently demonstrated by Chomaz and Gulminelli that bimodality 
of the probability distribution of the order parameter is equivalent
to the other definitions of phase transition proposed up to 
now~\cite{Chom03}.

\paragraph{Relationship to the Yang-Lee theorem}
The Yang-Lee theorem~\cite{Yang52} is considered as the standard definition 
of first-order phase transitions at the thermodynamic limit. As 
demonstrated 
in~\cite{Chom03} bimodality is a necessary and sufficient condition for 
zeroes of the partition sum in the control intensive variable complex 
plane to be distributed on a line perpendicular to the real axis. 

\paragraph{Anomaly of thermodynamical potentials}
A first order phase transition is characterized by an \textit{inverted 
curvature} of the relevant thermodynamical potential 
(entropy, free energy)~\cite{Chom04,Gros02}. 
This feature is also equivalent to a bimodality in the event 
probability of the given order parameter $X$ as displayed in 
the left part of fig.~\ref{otherdef}.

\paragraph{Negative derivatives of the thermodynamical potentials}
A first order phase transition was also related to a back-bending in 
the equation of state of the system~\cite{Chom04}, characterized by 
a negative second derivative of the thermodynamical potential, 
as for example the heat capacity 
if the energy is the order parameter, fig.~\ref{otherdef} right.

\subsection{Microcanonical vs. canonical ensemble}

\begin{figure}[h!]
\includegraphics[width=\columnwidth]{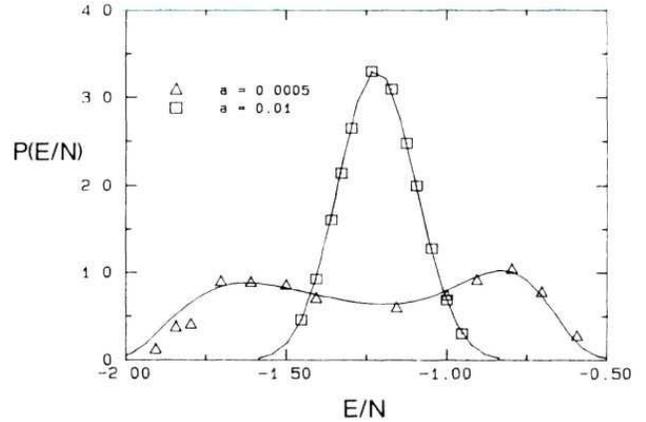}
\caption{Energy distributions obtained in the Gaussian ensemble
for different $a=N/N'$. Taken from~\cite{Chal88}.} \label{gauss}
\end{figure}
Among the observables signing a phase transition, 
the heat capacity is related to the fluctuations of the partial 
energy of the system and need to be studied in the microcanonical 
ensemble, while bimodality can only be observed when 
the system is free to fluctuate in terms of the associated extensive 
variable (i.e. energy or volume). This case corresponds to canonical
or isobar ensembles. In other words, events must be selected
 without constraint on the extensive variable in order to study bimodality. 
However in nuclear physics experiments, the two colliding nuclei 
form an isolated system: it seems thus natural to work in a 
microcanonical ensemble, and cuts can be applied on the energy of 
the system,
determined for instance by calorimetry. It seems conversely out
of reach to be in a canonical framework which would require the 
existence of a large heat bath.

The situation is not hopeless, as it was shown some years ago that
properties of phase transitions can be observed even if the working
ensemble is not strictly microcanonical or canonical, but is an
interpolating ensemble. In  Gaussian ensembles for instance, it
is supposed  that $N$ particles are in contact with a
system of $N'$ particles acting as a heat bath at temperature T.
When $N'$ varies from 0 to $\infty$, the working ensemble mimicks
the transition between microcanonical and canonical~\cite{Chal88}.
fig.~\ref{gauss} presents the results of such a simulation, where it is
clearly  seen that the probability distribution of the energy
- in the transition region - presents a bimodal shape only when $N/N'$
is small enough ($<1/1000$) while for larger $N/N'$, the situation 
is that of the 
microcanonical case with only one peak in the distribution. 

\subsection{Liquid-Gas phase transition}
\begin{figure}[h!]
\centering
\includegraphics[width=0.9\columnwidth]{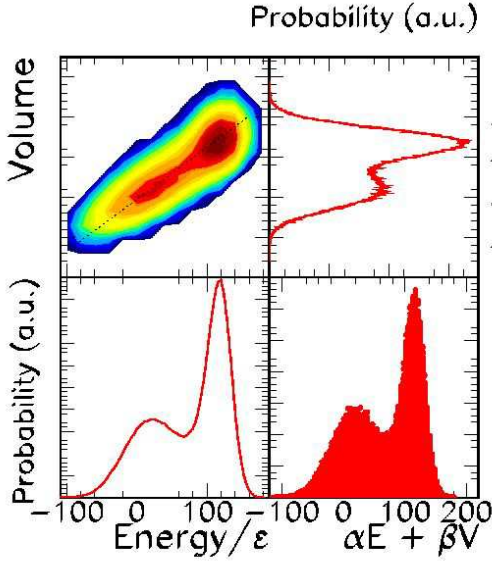}
\caption{Probability distributions of the energy E, volume V and a
combination of the two variables coming from a Lattice-Gas simulation 
in the canonical ensemble.
Taken from~\cite{Chom01}.} \label{EV}
\end{figure}
Since nuclei are supposed to undergo a liquid-gas phase transition, 
specific studies of this peculiar transition were undertaken through
Lattice-Gas calculations. 
In liquid-gas phase transitions, volume as well as energy are order 
parameters. The bimodality of the event probability distribution  in 
the first order phase transition region is evident in fig.~\ref{EV} 
which shows the location of events in the Volume \textit{vs.} 
Energy plane (top left). The projections along the axes ($E,V$) 
also display the expected bimodality, as does a linear combination 
of these two order parameters (bottom right, red curve). In this 
framework (Lattice-Gas model), bimodality is evidenced if we are 
able to select (sort) events in a canonical way (or as close as possible, 
see previous section), and plot the event probability distribution of the 
energy or volume, or any observable directly related to them.

\section{Experimental observations}

Since bimodality was proposed as a signature of liquid-gas phase 
transition, it was extensively searched for in event samples resulting 
from nuclear collisions; studies were made for central collisions, where 
the liquid-gas phase transition is clearly evidenced by previous analyses 
(see chapter ``phase transition'') as well as for peripheral collisions, 
where a large range of excitation energy can be explored.

\subsection{Central collisions: systems with mass $\sim$ 250}
\begin{figure*}[htb]
\centerline{\includegraphics[width=0.8\textwidth]{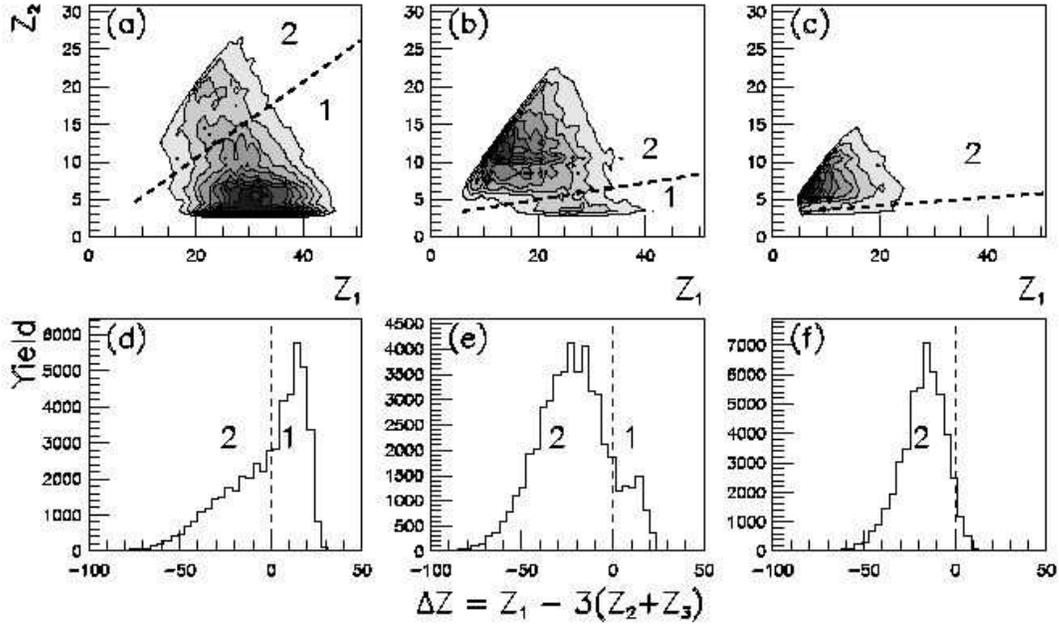}}
\caption{Correlation between the two largest fragments, $Z_1$ and
$Z_2$ obtained in  central collisions for the Ni+Au system at
32$A$ (left), 52$A$ (middle) and 90$A \,$MeV (right). 
The bottom row shows an asymmetry variable
 built as a linear combination of the atomic number of the three largest 
 fragments.
 Taken from~\cite{Bell02}.}\label{NiAu}
\end{figure*}
Systems with total mass close to 250 were studied with the INDRA array 
using two entrance channels, an asymmetric one, Ni+Au, and an almost 
symmetric one, Xe+Sn. In both cases, in the incident energy range 
scanned, it was shown that a fused system was formed in central 
collisions.
 
Bellaize et al~\cite{Bell02} have reported the observation of bimodality 
of the size asymmetry of the two largest fragments in central events 
for the Ni+Au system at $32A, \; 52A$ and $90A \,$MeV. It was
associated with two fragmentation patterns (see first row of 
fig.~\ref{NiAu}), one similar to residue-evaporation (one large fragment 
with few small ones,  zone~1 in fig.~\ref{NiAu}), the other to 
multifragmentation (fragments of nearly equal size, zone~2). 
A variable built with the charges of the three largest fragments,
$Z_1, Z_2, Z_3$ in decreasing order,
\begin{equation}
Z_1 - 3(Z_2 + Z_3) \label{asyOL}
\end{equation}
also has 
a bimodal distribution at $32A$ and $52A \,$MeV, as shown in the 
bottom row of fig.~\ref{NiAu}, but no longer at $90A \,$MeV. 
This fact is compatible with the location of the system in 
the coexistence region below $52A \,$MeV, where it can experience a first 
order phase transition by exploring different densities and temperatures. 
For higher energies (here $90A \,$MeV), the system passes directly 
through the coexistence region and we observe only the presence of 
the multifragmentation regime, which could indicate that the 
system explores only the low density part of the phase diagram.

\begin{figure}
\includegraphics[trim= 0 0 0 66,clip,width=0.9\columnwidth]{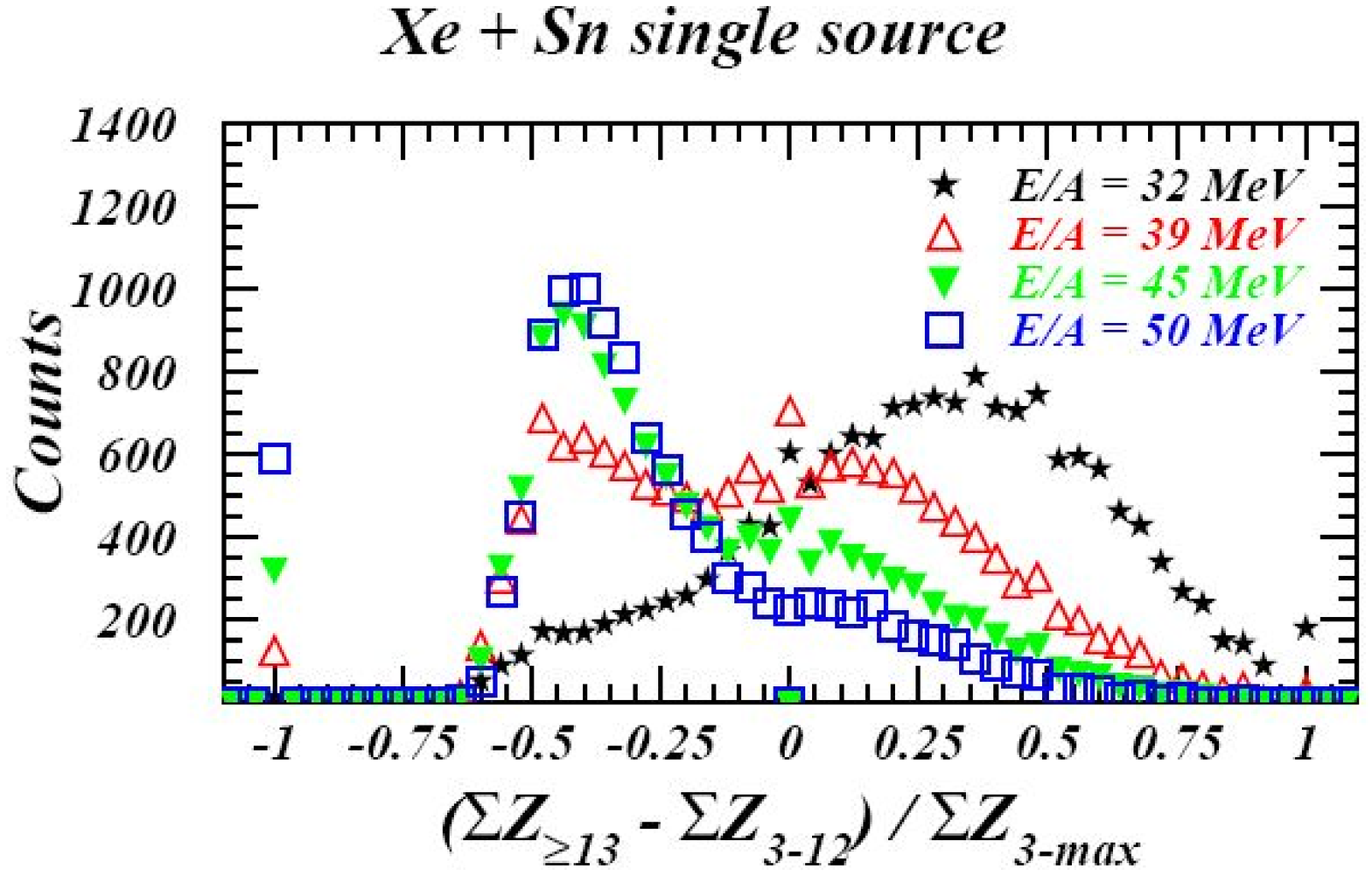} 
\caption{Probability distributions of the charge asymmetry between light
(Z=3-12) and heavy fragments (Z$\geq$12) for fused events in the Xe+Sn system 
at 32$A$, 39$A$, 45$A$ and 50$A \,$MeV. 
Taken from~\cite{Bord02}.} \label{XeSn}
\end{figure}
fig.~\ref{XeSn} shows the distributions obtained when looking at the
asymmetry ratio between heavy, ($Z \geq 13)$, and light, (Z=$3-12$),
fragments
\begin{equation}
\left( \sum{Z_{\geq13}} - \sum{Z_{3-12}} \right) /
\sum{Z_{\geq3}} \label{asyBB}
\end{equation}
\begin{figure}
\includegraphics[width=0.9\columnwidth]{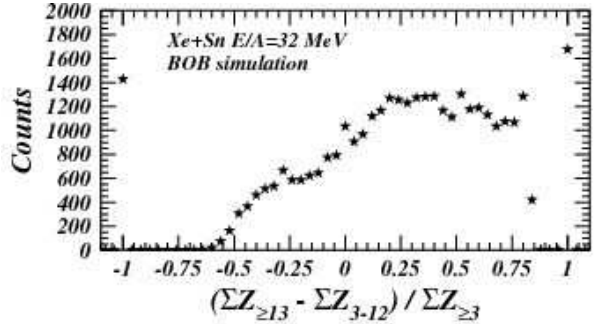}
\caption{Charge asymmetry obtained by using a Stochastic Mean-Field simulation 
(BOB\cite{Chom91}) for central events of the Xe+Sn system at 32A \,MeV. 
Unpublished results from the authors of~\cite{Tab03}.} \label{BOB}
\end{figure}
 for single source events produced in central
Xe+Sn collisions between 32$A$ and 50$A\,$MeV~\cite{Bord02}.
Bimodality is present at all energies, with dominant ``liquid-type'' 
events at \\ $32A\,$MeV,
and a dominance of ``gas-like'' events at and 
above $45A\,$MeV; the two types of events are in roughly equal number
at $39A\,$MeV, where other phase transition signals have been already 
observed (see chapter ``Many fragment correlations''). The authors
of~\cite{Bord02} relate the chosen asymmetry variable to the density 
difference between the coexisting liquid and gas phases of 
nuclear matter. 
The same variable was built for the events resulting from a 
Stochastic Mean Field simulation \cite{Chom91} of head-on collisions between 
Xe and Sn at 32$A\,$MeV.  In this simulation, which was shown to 
well reproduce many experimental features, single variable distributions 
as well as different correlations~\cite{Fra01,Tab03,Tab05} 
(see chapter ``Many fragment correlations''), the system enters the 
coexistence region and multifragments through spinodal decomposition.
The equivalent of fig.~\ref{XeSn} for simulated events is shown in 
fig.~\ref{BOB}; the picture is very similar to the experimental data 
at the same energy (black stars in fig.~\ref{XeSn}), a bimodal behaviour 
appears with a dominance of events of liquid type.

\subsection{Central collisions: systems with mass $\sim$ 100}
\begin{figure}
\centering
\includegraphics[width=0.9\columnwidth]{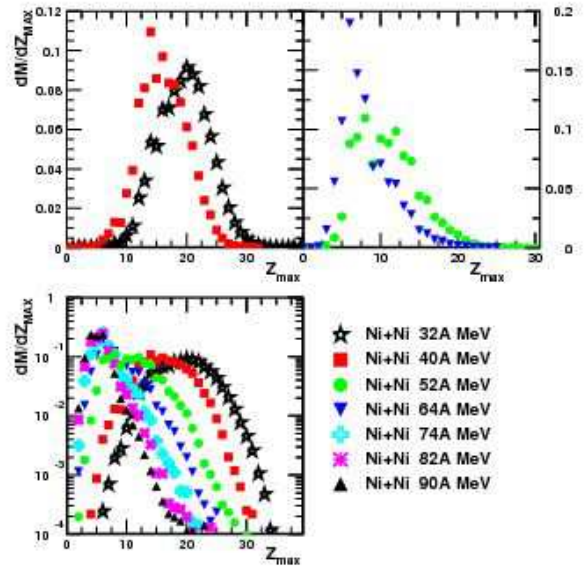}
\caption{Distributions of the largest fragment for central Ni+Ni collisions 
from 32 to 90$A \,$MeV (bottom). The same distributions at the four lowest
energies are displayed in linear scale in the top panels. 
Taken from~\cite{Lau06}.}\label{NiNi}
\end{figure}
Central collisions between two $^{58}$Ni nuclei were studied at incident 
energies between 32 and 90$A\,$MeV; event selection was made through a 
discriminant factorial analysis trained, at variance with ref~\cite{Lau05}, 
on the complete experimental events. A bimodal distribution of the largest
fragment was observed at 52$A \,$MeV, intermediate between the Gaussian
distributions measured at lower energies and the asymmetric distributions 
found from 74$A\,$MeV up~\cite{Lau06}.
The minimum is rather shallow (about 80\% of the peak value); at 
64$A\,$MeV a bimodal distribution persists, but now the
peak on the more fragmented side is dominant. Conversely the distributions
of the fragments of higher rank (not shown) are monotonous. To our knowledge, 
it is the only direct observation of bimodality on the largest fragment.

\subsubsection{Going further}

Central collisions allowed to study and evidence a bimodal behaviour
of some asymmetry variables, which can be connected to the density 
difference between a liquid and a gas phase; in that sense 
they would be good candidates for being order parameters of a 
liquid-gas type transition. 
Nevertheless, several drawbacks can be pointed out; firstly it was 
shown that the lighter fragments exhibit a pre-equilibrium component 
in Ni+Au~\cite{Bell02}, while radial flow effects were recognised in 
symmetric systems, Xe+Sn~\cite{Mar97,NLN99,Bor04} and Ni+Ni~\cite{Lau06}. 
But above all, the sorting of central events 
selects a rather narrow region in excitation energy for each incident
energy (about 1-2$A\,$MeV at half maximum of the distribution).
This is closer to a microcanonical working ensemble and
may prevent a very clear observation of bimodality.

\subsection{Quasi-projectiles in peripheral collisions}

Analyses of quasi-projectiles formed in peripheral and 
semi-peripheral reactions are thus mandatory, as they allow
to overcome some of the abovementioned problems. In particular
a broad excitation energy distribution of quasi-projectiles (QP)
can be accessed. Exchanges of energy and particles 
with the quasi-target (QT), while it lies in the neighbourhood of the QP
and especially when it is heavy, mimick a small heat bath  and 
an almost canonical sorting can be envisaged. 
Whenever the incident energy is high enough, the different components 
(the QT and the QP, and the pre-equilibrium or neck part)
can be better disentangled, or at least the uncertainties caused 
by their existence can be circumvented.

Most of the studies on quasi-projectiles arise from Au on Au
collisions at various energies. 
Extensive results concerning a very light nucleus, close to Argon
were also recently proposed. Several variables are used for sorting 
events as a function of
the violence of the collisions; among the most commonly employed one 
can cite multiplicities and the transverse energy (relative to 
the beam axis) of charged  products, either all of them or only light 
charged particles ($Z=1,2)$~\cite{Llo95,Pete95,Luka97,Fra001}. 
Other sorting are based on Z$_{bound}$ (the sum of charges for fragments, 
$Z>2$), as proposed by the ALADIN collaboration~\cite{Sch96}, 
or on the excitation energy (NIMROD collaboration)~\cite{Ma04}. 

\subsubsection{Au quasi-projectiles at relativistic energies}
\begin{figure}[hbt]
\centering
\includegraphics[width=0.8\columnwidth]{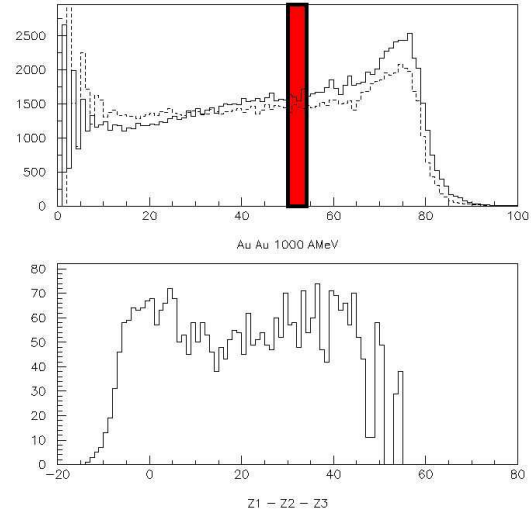}
\caption{Z$_{bound}$ (top) and charge asymmetry distributions (bottom)
 for the Au+Au system at 1A GeV. The bottom panel corresponds to the 
Z$_{bound}$ selection displayed by the highlighted area in the top panel.
 Taken from~\cite{Trau05}.} \label{aladin}
\end{figure}
\begin{figure*}[hbt]
\centering
\includegraphics[scale=0.5]{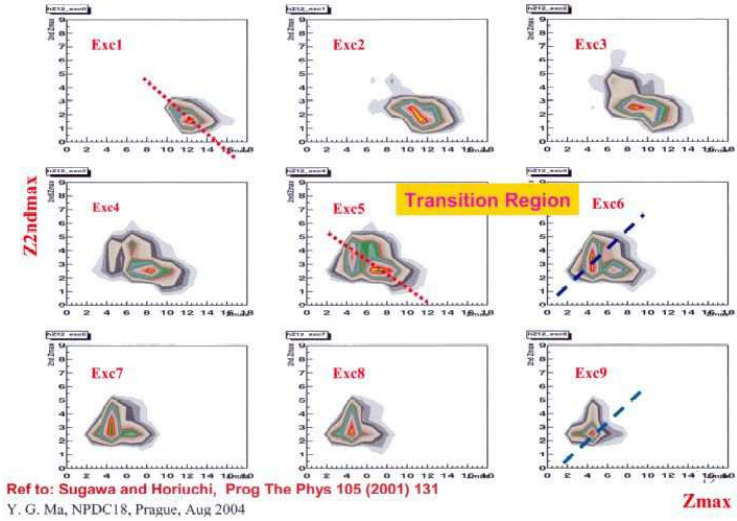}
\caption{Correlation between the largest charge (Z$_{max}$)
and the second largest (Z$_{2ndmax}$) for the QP in peripheral Ar+Ni
collisions at 47$A \,$MeV. Panels from Exc1 to Exc9 correspond to a 
selection in increasing excitation energy (see text). 
Taken from~\cite{Ma04}.} \label{Ma}
\end{figure*}
The ALADIN collaboration reported the presence of bimodality for 
peripheral Au+Au reactions at $1A$ GeV~\cite{Trau05}. 
Fig.~\ref{aladin} shows the $Z_{bound }$ distribution (top panel) 
where is highlighted the selected region, $Z_{bound}=53-55$, 
for which was drawn the charge asymmetry between the three 
largest fragments
\begin{equation}
Z_1-Z_2-Z_3 \label{asyWT}
\end{equation}
in the bottom panel. 
The charge asymmetry exhibits two components, the first one 
centered at low values (close to 0), which is associated to 
multifragmentation events, and the second one located at values 
around 40, which is more likely due to an evaporation residue of 
charge $Z$ close to $Z_{bound}$. It is worth saying that a 
percolation simulation was able to reproduce this bimodality in the 
charge asymmetry at the transition point. In this case, this is a second 
order phase transition. This point will be discussed in the 
``pending questions`` section.

\subsubsection{A smaller system with mass $\sim$ 40}
In a very complete analysis, Ma~\emph{et al}~\cite{Ma04} 
scrutinized data collected with the NIMROD array. They were able to 
reconstruct, from their emitted particles and fragments, the 
quasi-projectiles formed in  47$A\,$MeV Ar+Al,Ti,Ni %semi-peripheral 
collisions. 
The method used consisted in tagging the particles with
the help of a three moving source fit (QP, QT and mid-rapidity)
and then attributing to each of them, event per event, a probability
to be emitted by one of these sources. Completeness of quasi-projectiles,
($Z_{QP} \geq$12), from  semi-pe\-ri\-pheral collisions was further 
required; QP excitation energy was determined using the energy balance 
equation. The distributions of excitation energy so obtained for the 
three targets superimpose, showing that the QP excitation 
energy calculation is under control.

Plots of the charge of the second largest fragment \textit{vs.} the
largest one are shown in fig.~\ref{Ma}. As for heavier systems,
the topology evolves from residue-evaporation to multifragmentation
with increasing excitation energy. An equipartition of events
between two topologies is observed for $E^*/A=$5.5~MeV, where
at the same time fluctuations on the size of the largest
fragment are the largest, the power-law exponent 
for the charge distribution is minimum, and scaling laws are present. 
Here again, bimodality is observed at the same time as other possible
indicators of a phase transition.

\subsubsection{Toward a canonical event sorting?}

In the previous cases the sorting for peripheral reactions uses
properties of the studied source itself (here the QP) and is then 
probably more akin to a microcanonical than a canonical sorting.
Indeed the bimodal character of the distribution is not
very marked, as expected if the experimental sorting constrains strongly
the excitation energy~\cite{Gul05}.
To attempt a true canonical sorting one must discriminate the studied 
system from some heat bath. A first tentative in that aim was the 
study of Au quasi-projectiles through a sorting performed on
the transverse energy of the particles of the Au quasi-target (as the
system is symmetric, this amounts to particles emitted backward in 
the c.m.). This sorting is illustrated by the results presented hereafter.

\subsubsection{Au-like nuclei in a "canonical" sorting}
\begin{figure*}[htb]
\begin{minipage}[t]{0.95\columnwidth}
\includegraphics[width=\columnwidth]{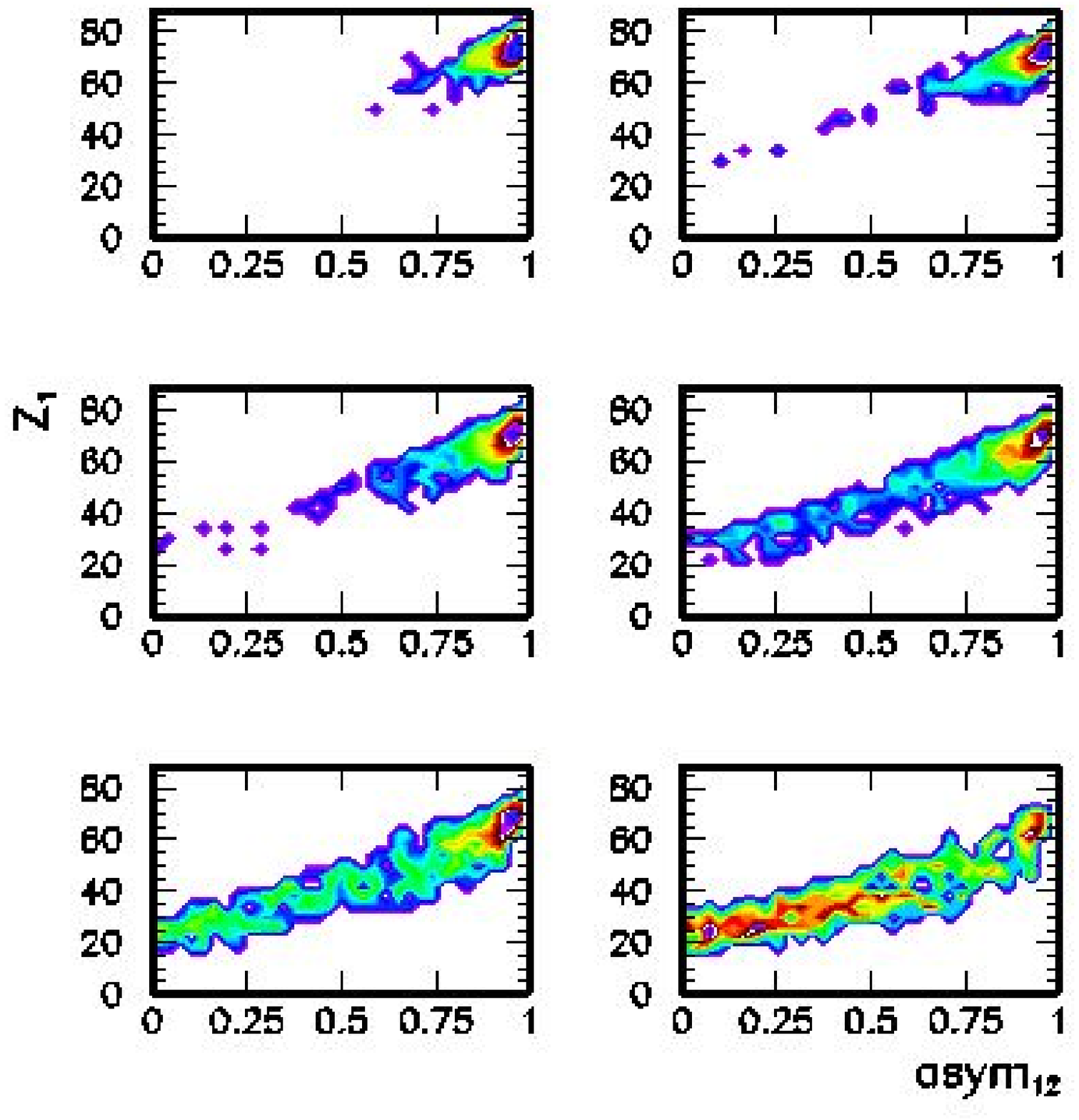}
\caption{Correlation between the charge of the largest fragment 
(Z$_{1}$) and the charge asymmetry (asym$_{12}$) 
between the two largest fragments for peripheral
 events of the Au+Au system at 35$A\,$MeV.
The panels correspond to a selection in 
increasing transverse energy of particles coming from the QT side
from top left to bottom right. Taken from~\cite{Dago04}.}
\label{AuAu1}
\end{minipage}%
\hspace*{1.5\columnsep}
\begin{minipage}[t]{0.95\columnwidth}
\includegraphics[width=\columnwidth]{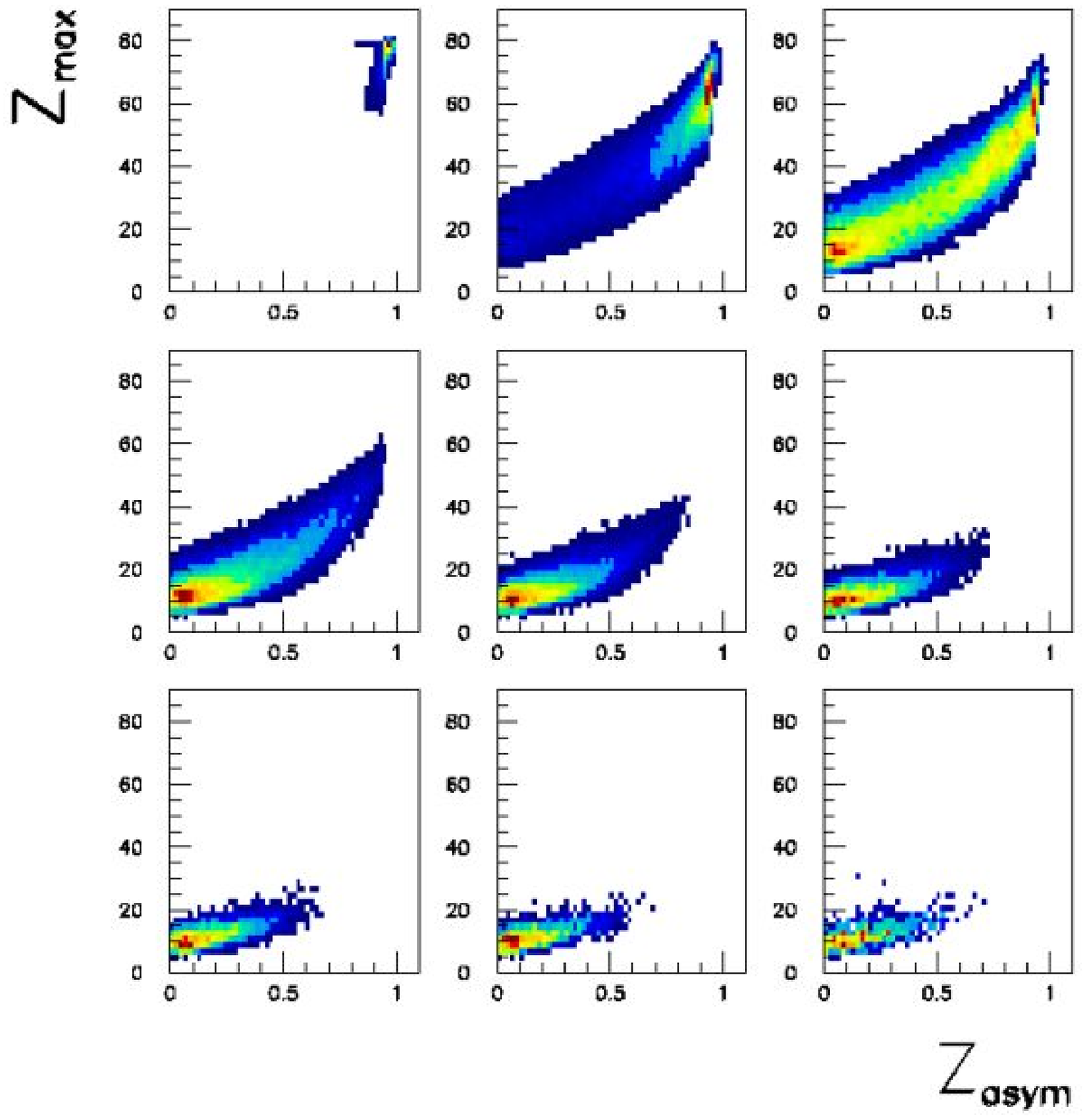}
\caption{Same as for fig.~\ref{AuAu1} but for Au+Au at 80$A\,$MeV. 
The panels corresponds to the same cuts in QT transverse energy.
Taken from~\cite{Pich04}.}
\label{AuAu2}
\end{minipage}%
\end{figure*}
Au quasi-projectiles from Au+Au collisions at various incident energies
were widely studied. Two examples are given here, at $35A\,$MeV 
-~results from the MULTICS-MINIBALL collaboration~\cite{Dago04}~- 
and at $80A\,$MeV, data from the INDRA/ALADIN collaboration~\cite{Pich04}. 
In both cases data were sorted versus the transverse energy of the QT 
light charged particles. The charge of the largest fragment 
in each event is plotted in figs~\ref{AuAu1},\ref{AuAu2} versus 
the charge asymmetry of the two largest fragments, 
\begin{equation}
(Z_1 - Z_2)/(Z_1+Z_2). \label{asyBT}
\end{equation}
Whatever the incident energy the picture evolves from 
an evaporation residue to a multifragmentation configuration, 
passing through a zone where the two topologies coexist, separated 
by a neat minimum; in this zone (last one at $35A\,$MeV, third one 
at $80A\,$MeV) 
the distributions present a bimodal behaviour. Note that the bimodal
character is not very strong when one projects the bidimensional
figures on either Z$_{max}$ or on the asymmetry. This is attributed 
in~\cite{Pich04} to the presence of preequilibrium effects, and some 
remaining aligned momentum which tend to shallow the minimum of a
bimodal distribution.

\section{Bimodality in models}

Different statistical as well as dynamical models explicitely or 
implicitely contain a phase transition. They predict the 
occurrence of bimodal distributions for selected variables around 
some transition energy. Examples are given in this section.

\subsection{\textit{SMM}: Statistical Multifragmentation Model}
\begin{figure}[h!]
\centering
\includegraphics[width=0.8\columnwidth]{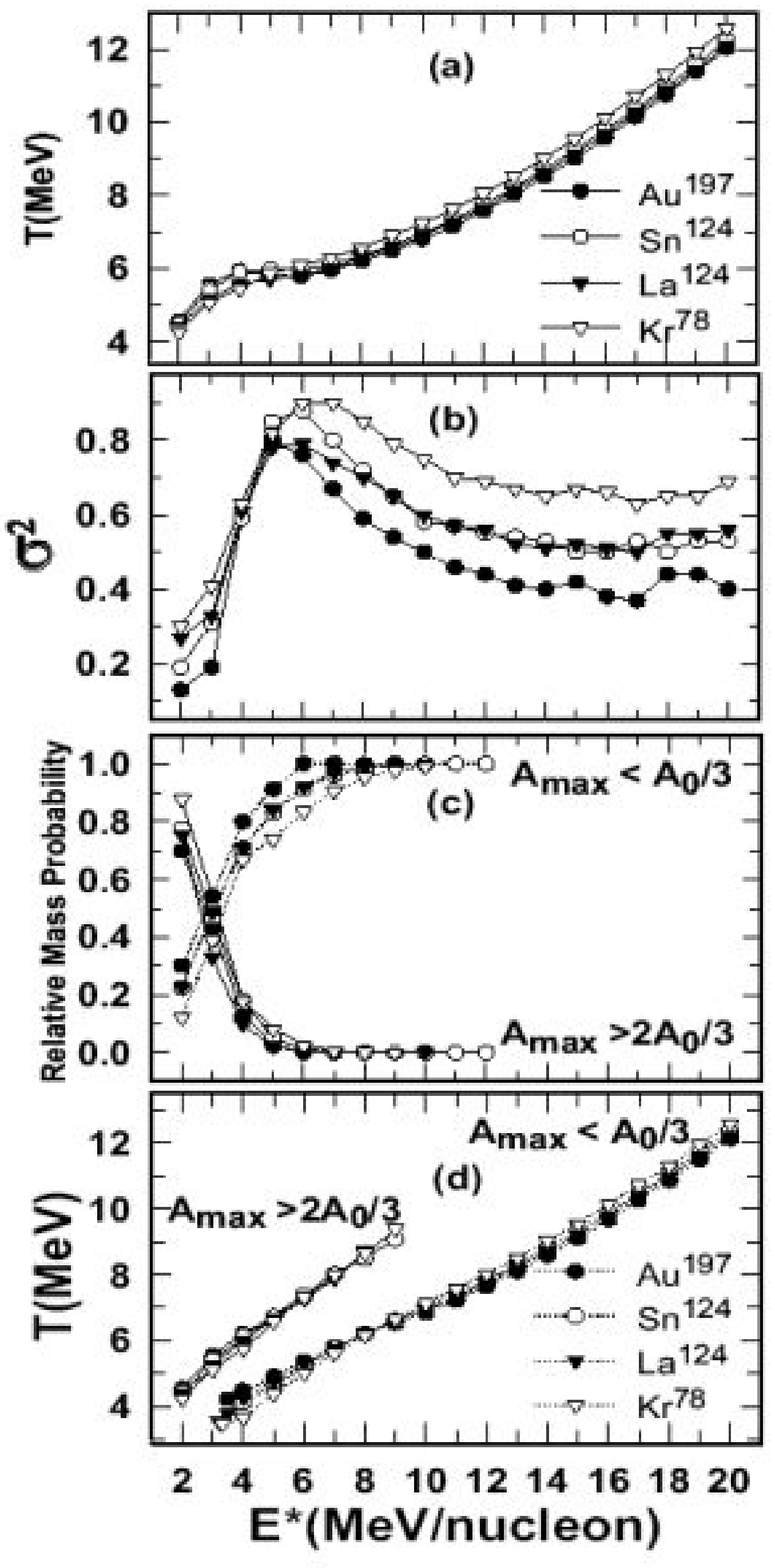}
\caption{Temperature average value (a) and variance (b), probability of
events selected on A$_{max}$ (c) and average temperature for these
events (d) versus $E^*$ for Kr, La, Sn and Au nuclei 
(\textit{SMM} simulations). Taken from~\cite{Buyu05}.} \label{SMM}
\end{figure}
Buyukcizmeci, Ogul and Botvina~\cite{Buyu05} analyzed \textit{SMM} 
simulations for heavy nuclei of various sizes, with excitation 
energy ranging from 2 to 20~MeV/nucleon. They found that all nuclei 
exhibit the same caloric 
curve, depicted in the top panel of fig.~\ref{SMM}, with the well-known 
``plateau'' between 4 and 7~MeV/nucleon (note in passing that the common
temperature at plateau whatever the mass of the considered nucleus is
in contradiction with the experimental results analyzed in 
ref~\cite{Nat02}). In the same energy interval as the plateau, 
the fluctuations of $A_{max}$ (not shown) and of the temperature 
(panel (b) of the figure) are maximum.
The authors sorted the events following the size of the largest
fragment, $A_{max}$. They defined two event classes, one with
$A_{max}\geq 2A_0/3$, representative of the residue-evaporation 
channel and the other one with $A_{max} \leq A_0 /3$, characterizing
multifragmentation events - $A_0$ being the total system size.
Panel (c) of fig.~\ref{SMM} shows that the probability of the first 
group decreases rapidly in the excitation energy range 
2-6~MeV/nucleon, while
that of the second one increases. The temperatures $T$ associated to
each class are different, as appears on the related caloric curves:
the residue-evaporation class shows a Fermi-gas behaviour 
(proportional to $T$-squared) while the multifragmentation class is 
associated to a classical gas (linear in $T$). The combination of 
these two behaviours gives rise to the plateau zone in the total 
caloric curve and explains the inflexion point of this curve. 
One is thus dealing with a direct bimodal behaviour, 
with two excitation energies associated with one
temperature in the transition region. This behaviour is an intrinsic 
feature of the phase space population in the \textit{SMM}.

\subsection{\textit{CMD}: Classical Molecular Dynamics}

\begin{figure}[h!]
\includegraphics[width=\columnwidth]{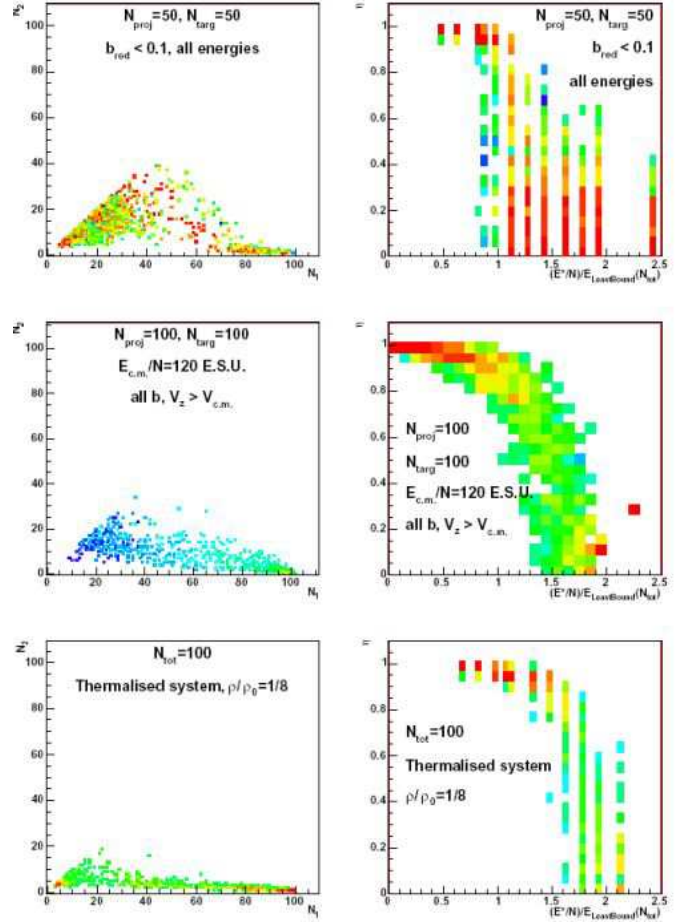}
\caption{Mass correlation between the two largest fragments (left)
and mass asymmetry, $\eta$, as a function of the excitation
energy (right), for collisions of LJ droplets. The top panel 
is associated to central  collisions; the middle to peripheral ones 
and the bottom panel to "thermalized" systems (see text)~\cite{Cus05}.}
\label{CMD}
\end{figure}

Signals of phase transition were searched for in dynamical
models. A simple example is a classical molecular dynamics
model with a Lennard-Jones potential implemented by Cussol~\cite{Cuss02}.
With such a potential, analogous to the van der Waals interaction for
fluids, the model includes a liquid-gas phase transition.
Symmetric collisions of \textit{LJ} droplets with sizes of $50+50$
and $100+100$ are analyzed.
Systems were prepared in three different conditions: 
\begin{itemize}
\item central collisions (small impact parameters), 
\item peripheral collisions (all impact parameters but looking at 
the forward zone, "quasi-projectiles"), 
\item "thermalized" systems (particles are placed
in a box of volume $V/V_0=8$ and released after a
time sufficient to reach thermal equilibrium). 
\end{itemize}
Two variables were scrutinized, the size asymmetry between 
the two largest fragments, $\eta$~(eq.~\ref{asyBT}),
and the mass correlation between these same fragments~\cite{Cus05}. 
The results, for systems comprising 100 droplets, are presented 
in fig.~\ref{CMD} (top: central collisions, middle: quasi-projectiles
and bottom: thermalized system). Excitation energies are expressed 
in ESU,  ratio between the excitation energy per particle and
the binding energy of the least bound particle.

Bimodality -~the occurrence of two fragmentation patterns in a given
energy zone~- is present in all situations, but at different 
excitation energies: $1~ESU$ for central collisions, $1.5~ESU$ for
quasi-projectiles, and $\sim1.8~ESU$ for the thermalized system. 
It is however worth to mention that if the thermalized systems 
is prepared at higher densities ($\rho $/$\rho_0=1-1.5$) 
the transition between the fragmentation patterns also
occurs but at lower excitation energy, namely $<1~ESU$~\cite{Cus05}. 
Cussol attributes the differences in the transition energy to the 
lack of complete thermalization of any source produced in nuclear 
collisions, whatever the impact parameter. 
One can conversely argue that this study proves
that bimodality is a robust signature of phase transition, as it 
survives even if the system is not fully thermalized,
although the apparent transition energy is displaced. This point  
will be developed later.

\subsection{\textit{HIPSE}: Heavy Ion Phase Space Exploration}

The Heavy Ion Phase Space Exploration model \textit{HIPSE} comprises a full 
(classical) treatment of the entrance channel (nucleus-nucleus potential, 
\textit{NN} collisions). It is followed by a random sampling of nucleons 
in the participant zone from Thomas-Fermi distributions of the two 
colliding nuclei to form fragments in the dense zone~\cite{Lauw04}. 
Excitation energy is shared among all products, taking into account 
the total energy constraint. Finally a statistical de-excitation 
(\textit{SIMON} code~\cite{Dura92}) of the fragments, including QP and QT 
- if they are still present - is performed.

\begin{figure}[h!]
\includegraphics[width=\columnwidth]{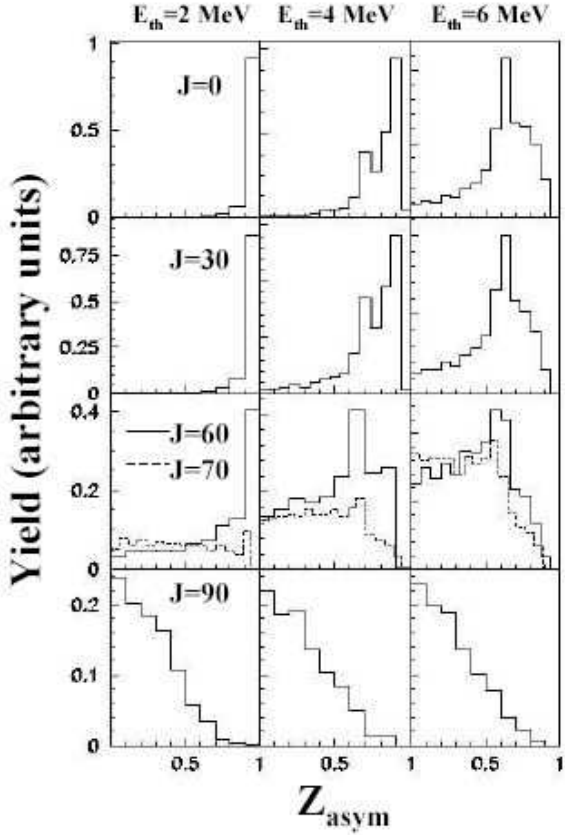}
\caption{Charge asymmetry distributions resulting from the de-excitation 
of hot Sn nuclei with different initial excitation energies 
(columns) and spin (rows), with the \textit{SIMON} code.
Taken from \cite{Lope05}.} \label{HIPSE}
\end{figure}
Simulations were done for all impact parameters, to mimick a
real $50A\,$MeV Xe+Sn experiment, then the same analysis 
as in~\cite{Pich04} was performed by Lopez et al~\cite{Lope05}; 
a bimodal structure was observed in the correlation between 
$Z_{max}$ and the charge asymmetry of the two largest 
fragments~(eq.~\ref{asyBT}). In a model however one can go further and 
track the origin of the bimodal behaviour: is it due to 
the entrance channel (dynamical effect) or to the de-excitation step? 
The first hypothesis was ruled out, as no discontinuity was found in the 
evolution of the size of the hot largest fragment with the impact 
parameter: the bimodality was clearly attributed to the statistical 
de-excitation of the QP. 
A deeper analysis of the de-excitation stage was then 
achieved through the simulated statistical de-excitation of Xenon 
nuclei of different excitation energies and spins with the \textit{SIMON} 
code~\cite{Dura92}. This is depicted in fig~\ref{HIPSE}, where the 
distributions of the asymmetry variable~(eq.~\ref{asyBT}) are 
plotted for several initial conditions.
Increasing the excitation energy does decrease the average charge 
asymmetry, but never down to the small values observed in the data. 
Conversely, if more spin is given to the nucleus, the asymmetry 
variable displays a sharp transition around $60-70\hbar$, 
which corresponds indeed to the angular momentum for which the 
symmetric fission barrier vanishes.

The authors of~\cite{Lope05} conclude that, in the \textit{HIPSE} model, 
the observed bimodality found its origin in the spin 
rather than in the excitation energy transferred to the QP, 
being still a phase transition but not of the liquid-gas type.
It is worth mentioning that using the \textit{SMM} model 
for the de-excitation stage, the authors also observe bimodality 
in the size of the largest fragment. This is not surprising in view 
of the abovementioned study with the \textit{SMM}. 
However, this raises the important issue - still under debate - 
of the order parameters (and then the type of the phase transition) 
which govern the bimodality. This point will be 
discussed in the perspectives.

\section{Pending questions}

As seen in the previous sections, bimodality is a very common feature 
in nuclear collisions at intermediate energies. It is present in central 
as well as in peripheral collisions. It takes place for a large range 
of masses, A=40-200. It was however mentioned in the course of 
the text that it is experimentally difficult to isolate a source, 
because of dynamical effects leading to a mixture of pre-equilibrium
products and of QP/QT de-excitation particles. Even if a source can be
properly defined, one has to verify its degree of thermalisation. Indeed 
radial flow was found, particularly in central collisions, and 
transparency effects were also evidenced~\cite{Esc05}. It seems however
from both experimental~\cite{Pich04} and theoretical~\cite{Lope05} studies 
that bimodality is not mainly driven by dynamical effects. Ambiguities
remain in the type of phase transition observed, and consequently on the 
definition of a true order parameter. Some of these questions were 
addressed recently and are presented in the following.

\begin{table}[h!]
\caption{World-wide experimental results on bimodality in July 2005.
.}
\label{tab1}
\begin{tabular}{|p{2cm}|c|c|c|}
\hline
\textbf{Results} & \textbf{Reaction} & \textbf{Source} & \textbf{Bimodal} \\
\textbf{from}    & \textbf{centrality} & \textbf{size} & \textbf{variable} \\
\hline
\textbf{INDRA} & Central & $\sim $ 200 & $Z_{1}-Z_{2}$ (eq.~\ref{asyOL},
\ref{asyBT})  \\
\hline
\textbf{INDRA} & Central & $\sim $ 200 & $Z_{liq}-Z_{gas}$ (eq.~\ref{asyBB}) \\
\hline
\textbf{INDRA} & Central & $\sim $ 100 & Z$_{max}$  \\
\hline
\textbf{INDRA} & Peripheral & 160-180 & $Z_{1}-Z_{2}$ (eq.~\ref{asyBT})  \\
\hline
\textbf{MULTICS/ MINIBALL} & Peripheral & $\sim $180 & $Z_{1}-Z_{2}$ 
(eq.~\ref{asyBT}) \\
\hline
\textbf{ALADIN} & Peripheral & $\sim $130 
& $Z_{1}-Z_{2}-Z_3$ (eq.~\ref{asyWT})  \\
\hline
\textbf{NIMROD} & Peripheral & 24-40 & $Z_{liq}-Z_{gas}$  \\
\hline
\end{tabular}
\end{table}
Table 1 gathers all experimental results on bimodality found so far. 
A glance at the table indicates that bimodality was essentially found 
in charge asymmetry variables comprising the two or three largest 
fragments of each event. Such variables can in some sense be related to
density difference between a dense (liquid) and a dilute (gas) phase; in
some models for instance the Fisher droplet model, the largest fragment
is assimilated to the liquid while all the other form the gas. 

\subsection{Are $Z_{max}$, $A_{max}$, or the asymmetry order parameters?}

\begin{figure}[h!]
\includegraphics[width=\columnwidth]{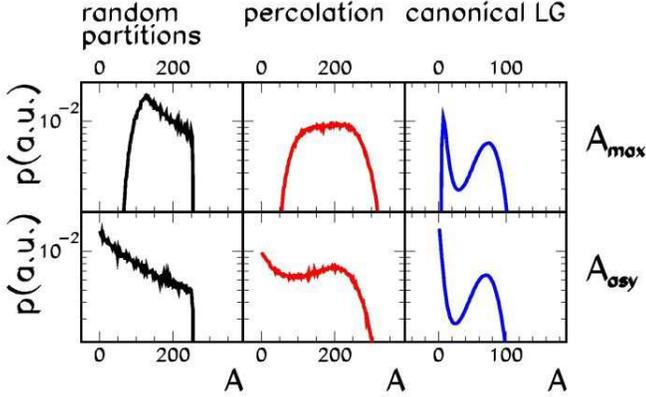}
\caption{Largest fragment A$_{max}$ (top) and mass asymmetry A$_{asy}$
(bottom) distributions for three simulations. Left: random partitions 
(no phase transition), middle: percolation (2$^{nd}$order phase transition) 
and right: canonical Lattice-Gas (1$^{st}$ order phase transition). 
Taken from~\cite{Gulm05}.} \label{PTorder}
\end{figure}

Simulations were performed in different frameworks to test whether 
the observables $Z_{max}$, $A_{max}$, or the asymmetry, reliably 
sign a phase transition. Let us recall
that a bimodality of an order parameter signs the occurrence of
a first order phase transition  in a finite system.
Fig.~\ref{PTorder} shows the outcomes of three simulations in 
the transition region - when it exists;
there is no phase transition in the Random Partitions calculation, 
while percolation has a second order transition and Lattice Gas a 
first order one~\cite{Gulm05}. The distributions of the largest fragment 
$A_{max}$  evidence that $A_{max}$ only presents
a bimodal distribution for the canonical Lattice-Gas calculation.
This means that $A_{max}$ is indeed an order parameter of  the
 first order  phase transition of the Lattice Gas.
 The distribution presents a wide plateau, as expected, 
 in the case of a continuous transition (percolation).
By contrast, the mass asymmetry, $A_{asy}$,  defined in a similar 
way as the charge asymmetry (eq.~\ref{asyBT}), also displays a bimodality 
(although with a less marked minimum) for 
simulations which have  a 2$^{nd}$ order phase transition
(percolation, middle column). The conclusion of this study is that both 
A$_{max}$ and A$_{asy}$ clearly signal a phase transition - note that 
none of them presents bimodality in a model without phase transition -
but A$_{max}$ is the only unambiguous signature of the order of the 
transition. 

\subsection{Order parameters of the liquid-gas phase transition}

If nuclei undergo a liquid-gas type phase transition, then the order
parameters are known: the energy, the volume. In some of the 
experimental studies cited above, the authors try to push the analysis 
beyond the single observation of bimodality on the asymmetry variable.
As a first attempt, in central collisions between Ni and Au at 
$52~A\,$~MeV~\cite{Bell02}, the excitation energies 
(experimentally deduced from the energy balance equation) associated 
to the two fragmentation patterns were found
slightly different (by $1 A \,$MeV)~\cite{Lope02}. This bimodality 
of the excitation energy is an indication in favour of the 
liquid-gas type of the phase transition observed.

\begin{figure}[h!]
\centering
\includegraphics[scale=0.6]{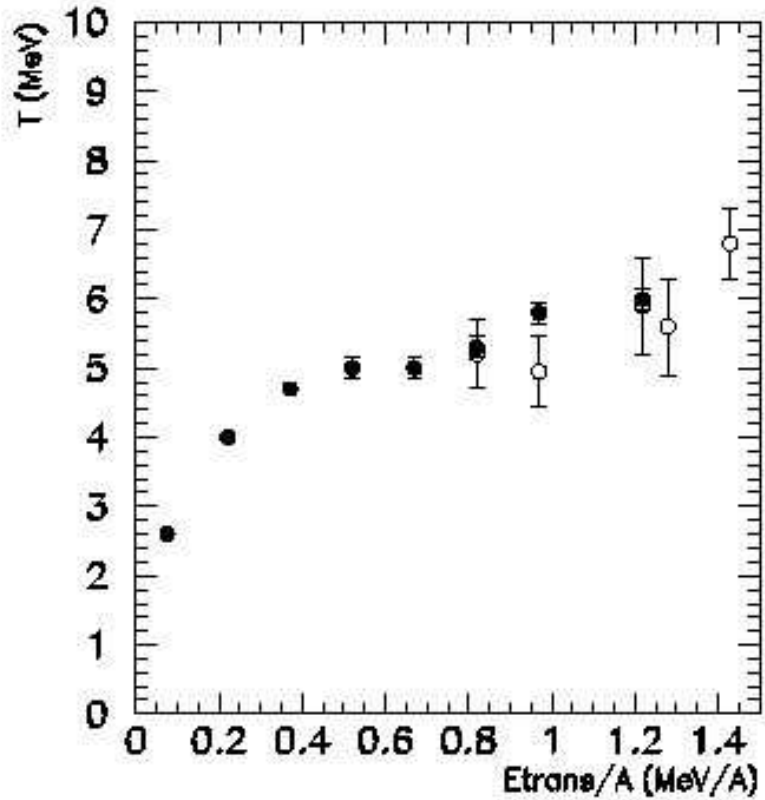}
\caption{Apparent temperatures of Au quasi-projectiles as a function of 
the normalized transverse energy for residue (filled symbols) and 
multifragmentation events  (open symbols). 
Taken from~\cite{Pich04}.} \label{temp}
\end{figure}
Studies of Au quasi-projectiles were deepened by the authors of 
ref.~\cite{Pich04}: a test of the reliability of the 
canonical picture was accomplished by estimating the apparent 
temperatures of the two types of events, from the slope of the 
emitted proton spectra  for residue-like events, 
and from double isostope ratios in the multifragmentation regime. 
As seen in fig.~\ref{temp} both temperatures are close enough in 
the region where bimodality is present  (Etrans~=~0.8-1.2$A \,$MeV),
while the excitation energies, calculated with the energy balance 
equation, are different. This is expected if bimodality has a thermal 
origin and validates the sorting as close to a canonical one. 

\subsection{Does bimodality survive out-of-equilibrium effects?}

\begin{figure}[h!]
\includegraphics[width=\columnwidth]{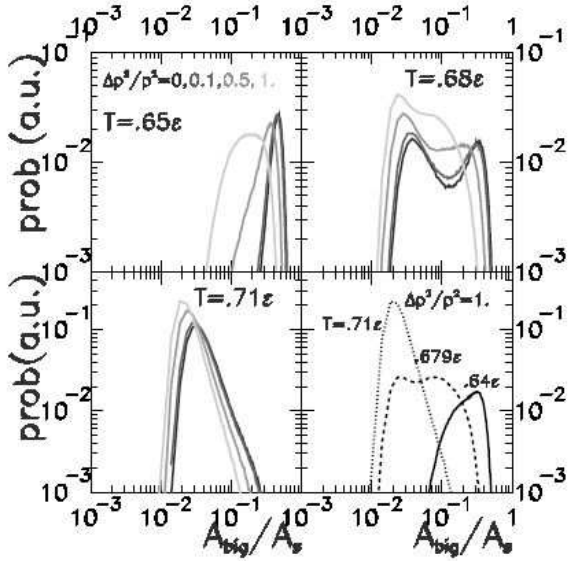}
\caption{Canonical Lattice-Gas simulations for different temperatures T 
around the critical one T$_{c}$=0.68$\varepsilon$ for the distributions 
of the largest fragment. Simulations are performed by adding an extra radial 
flow energy $\Delta$p$^{2}/p^{2}$ between 0 and 1. Taken from \cite{Chom01}.}
\label{outofeq}
\end{figure}

The influence of non equilibrium effects on signals of phase
transition was studied in~\cite{Gul04} in the case of incompletely
relaxed incoming momentum (transparency) and of self-similar radial
flow. Both effects were indeed recognized in experimental data.
Fig.~\ref{outofeq} displays  results of (canonical)
Lattice-Gas simulations with different radial flow energies; 

$\Delta p^2/p^2$ is equivalent to the ratio 
$\varepsilon_{flow}/\varepsilon_{th}$ and is varied from 
0 to 1. At the transition temperature, T=0.68$\varepsilon$,
a bimodality of $A_{max}$ is clearly seen in the absence of flow, 
and is still visible even when the flow energy is as important as 
the thermal energy (top right panel). The authors state thus that 
radial flow does disturb  the signal, partially filling the gap 
between the two components, but does not destroy it as long as the 
flow does not dominate the global energetics.
Similar conclusions were drawn in this paper in presence of 
longitudinal flow (transparency effects).

These two examples illustrate the robustness of  bimodality versus 
external (and realistic) constraints due to the dynamics of the 
collision; similar conclusions can also be derived from 
CMD simulations (see above).

In experimental data on Au quasi-projectiles~\cite{Pich04}, refined 
treatments aiming at better isolating quasi-projectiles from 
the mid-rapidity contribution and keeping only events where 
this contribution was smaller were tempted.
In all cases the bimodal picture comes out better, although it occurs
for a lower value of the sorting variable (smaller dissipation),
for a given incident energy. This is again an evidence of the 
robustness of bimodality against non-equilibrium effects.

\section{Perspectives}

Bimodality is a very promising signature of first order phase 
transition because of its simplicity and robustness against dynamical 
constraints. It was shown in this contribution that the signal is 
quite common for the decay of hot nuclei and can be observed in rather 
different experimental conditions (central / peripheral collisions, 
small / large source sizes). 

Nevertheless, some open questions need to be answered in order 
to firmly assess the validity of this signal. Several strategies
can be envisaged in order to progress in this direction:

\begin{itemize}
\item cross the observation of the bimodality signal with that of
all the other proposed
signals for the phase transition such as critical exponents,
scalings (Delta-scaling, Fisher scaling, Zipf law),
negative heat capacities, or space-time correlations (emission
times and correlation functions). Obviously, when possible,
all signals should be studied on the same sample of events to
minimise biases due to sorting. Such cross controls were started 
by the INDRA~\cite{Riv05} and NIMROD~\cite{Ma04} collaborations. 
One must solve however the problem 
of the non-equivalence of statistical ensembles in some cases.

\item test the effect of sorting. Indeed different ways of sorting were
proposed (impact parameter selectors, compact shape events, source 
selection). The robustness of any signal will be established if its
observation is not drastically dependent on the chosen sorting for
a given centrality for instance.

\item compare the results of different entrance channels for 
nuclear collisions; by using asymmetrical reactions such as 
light ions impinging on heavy targets, or nucleon/pion-nucleus 
reactions, one may hope to disentangle the
different effects which could possibly govern 
bimodality. By using these very different entrance channels 
reactions, the pre-equilibrium / neck contributions can be evaluated and 
even subtracted. Moreover, the effects of large collective motions 
such as radial flow (for central collisions) or spin (angular momentum 
transfer in semi-peripheral reactions) can also be measured. 
It will possibly help to answer to the fundamental question 
of the type of phase transition which is experienced by hot nuclei.
\end{itemize}

%%%%%%%%%%%%%%%%%%%%%%%%%%%%%%%%%%%%%%%%%%%%%%%%
%% BACKMATTER
%%%%%%%%%%%%%%%%%%%%%%%%%%%%%%%%%%%%%%%%%%%%%%%%
\begin{acknowledgement}
We thank all the nuclear physicists around the world who send us 
their results - published or not. 
\end{acknowledgement}

\end{document}